\documentclass[twocolumn,superscriptaddress,amsmath,aps]{revtex4}

\usepackage{amssymb,mathrsfs,bm,color,times}
\usepackage{graphicx,color}
\usepackage{CJK}
\usepackage{bm}
\usepackage[hypertex]{hyperref}
\usepackage{amsmath}

\newcommand{\be}{\begin{equation}}
\newcommand{\ee}{\end{equation}}
\newcommand{\bea}{\begin{eqnarray}}
\newcommand{\eea}{\end{eqnarray}}
\newcommand{\bsube}{\begin{subequations}}
\newcommand{\esube}{\end{subequations}}

\newcommand{\Eq}[1]{eq.\,(\ref{#1})}
\newcommand{\Eqs}[1]{eqs.\,(\ref{#1})}

\newcommand{\la}{\langle}
\newcommand{\ra}{\rangle}

\newcommand{\beq}{\begin{equation}}
\newcommand{\eeq}{\end{equation}}
\newcommand{\beqn}{\begin{eqnarray}}
\newcommand{\eeqn}{\end{eqnarray}}

\newcommand{\nl}{\nonumber \\}

\newcommand{\bsub}{\begin{subequations}}
\newcommand{\esub}{\end{subequations}}

\begin{document}
\begin{CJK*}{GBK}{Song}

\title{Probing nontrivial fusion of Majorana zero modes via near-adiabatic coupling}

\author{Jing Bai}
\affiliation{Center for Joint Quantum Studies and Department of Physics,
School of Science, \\ Tianjin University, Tianjin 300072, China}

\author{Luting Xu}
\affiliation{Center for Joint Quantum Studies and Department of Physics,
School of Science, \\ Tianjin University, Tianjin 300072, China}

\author{Wei Feng}
\thanks{Corresponding author: fwphy@tju.edu.cn}
\affiliation{Center for Joint Quantum Studies and Department of Physics,
School of Science, \\ Tianjin University, Tianjin 300072, China}

\author{Xin-Qi Li}
\thanks{Corresponding author: xinqi.li@imu.edu.cn}
\affiliation{Research Center for Quantum Physics and Technologies,
Inner Mongolia University, Hohhot 010021, China}

\affiliation{School of Physical Science and Technology,
Inner Mongolia University, Hohhot 010021, China}

\date{\today}

\begin{abstract}
{\flushleft We propose and simulate}
a near-adiabatically coupling probing scheme
for nontrivial fusion of a pair of Majorara zero modes (MZMs).
The scheme can avoid the complexity of oscillating charge occupation
in the probing quantum dot,
making thus practical measurements more feasible.
We also show how to extract the information of
nonadiabatic transition and fermion parity violation
caused during moving the MZMs together to fuse,
from the initial states prepared with definite fermion parity.
All the simulations, including the effective coupling between the fusing MZMs,
and their coupling to the probing quantum dot,
are based on the lattice model of a Rashba quantum wire
in proximity contact with an $s$-wave superconductor,
under the modulation of mini-gate voltage control.
\end{abstract}

\maketitle

{\flushleft\it Introduction}.---
In the past decade, considerable progress has been achieved
for realizing the MZMs in various experimental platforms \cite{DS23,Liu23,Gao22}.
Yet, the main experimental evidences are largely
restricted in the zero-bias conductance peaks,
which cannot ultimately confirm the realization of MZMs.
An essential next and milestone step is to identify the MZMs
by probing the underlying non-Abelian statistics,
via either braiding or fusion experiments.
Typically, the non-Abelian statistics of MZMs is reflected by quantum state evolution
in the manifold of highly degenerate ground states,
when braiding MZMs in real space \cite{Fish11,Opp12,Roy19,Han20}.
The fact that adiabatically braiding a pair of MZMs
would lead to a unitary evolution matrix,
rather than to a scalar phase factor,
indicates the quantum dimension $d>1$.
This can be equivalently reflected by fusing two MZMs from different pairs
to stochastically yield outcomes of either a vacuum,
or an unpaired fermion (resulting in an extra charge) \cite{Ali16,BNK20,Leij22,NC22,Li24a,Li24b}.
The nontrivial fusion with two outcomes can serve also
as a demonstration of non-Abelian statistics,
since it indicates the quantum dimension $d>1$.

\begin{figure}
  \centering
  \includegraphics[scale=0.35]{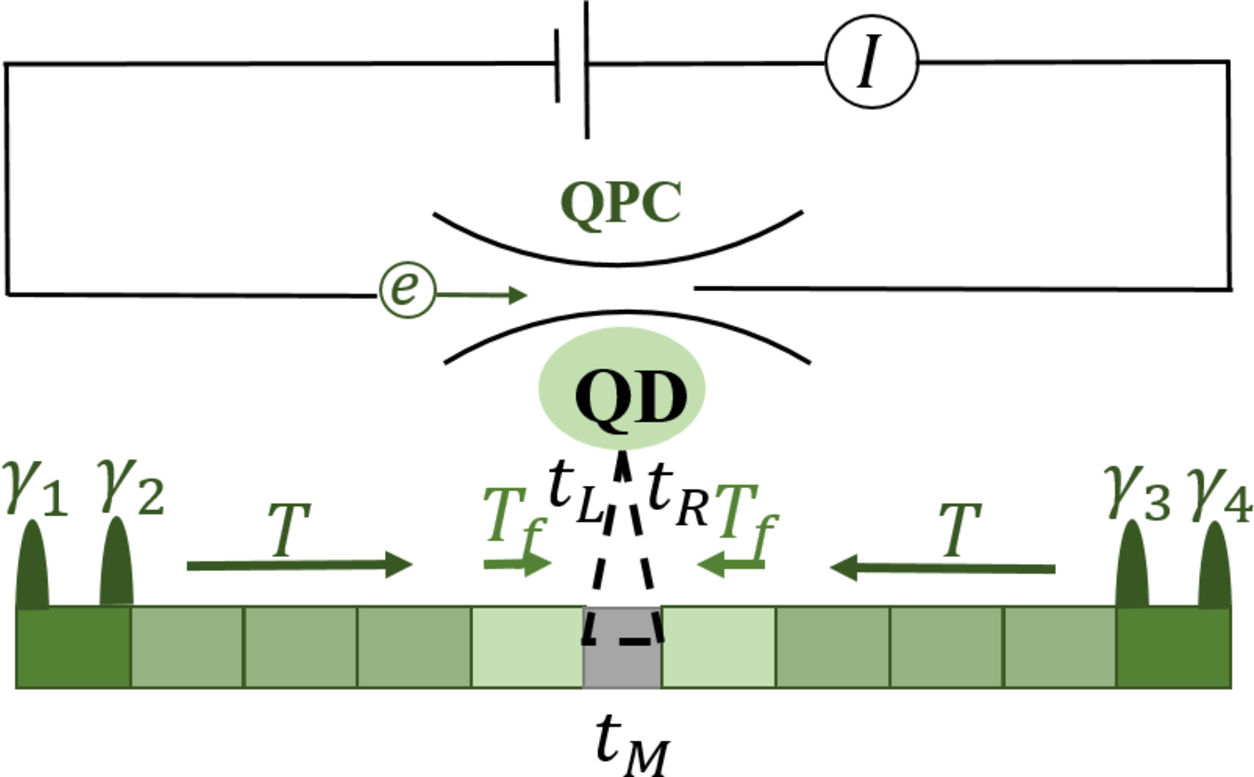}
  \caption{
Schematic diagram of mini-gate-controlled transporting
and fusing a pair of MZMs, $\gamma_2$ and $\gamma_3$,
from different Majorana pairs
prepared initially with definite fermion parity.
A single-level quantum dot (QD) is employed to probe the fusion outcomes,
while the charge change in the QD is measured
by a nearby quantum-point-contact (QPC) detector.
In this work, we propose to adiabatically switch on
the coupling of $\gamma_2$ and $\gamma_3$ to the QD,
when they are moved close to each other.    }
\end{figure}

In this work, following refs.\ \cite{NC22,Li24a,Li24b},
we perform simulations for the transport and fusion of MZMs,
as schematically shown in fig.\ 1,
starting with the initial state prepared with definite fermion parity.
The simulation is to base on the lattice model
of topological-superconductor (TSC) quantum wires,
under the modulation of mini-gate voltage control.
In principle, in order to
keep up the topological protection,
the transport of MZMs should be adiabatically slow.
However, in practice,
this may contradict other requirements,
such as avoiding the quasi-particle poisoning decoherence.
Thus, we will simulate the transport of a MZM with finite speed,
and pay particular attention to the effect of nonadiabatic transition
\cite{Ali18,Kel19,Byr21,Sar11,Opp13,Shn13,Karz14,Fran17,Sau19,Tru23}
and possible violation of the initial fermion parity
in a single wire (a closed system).
Importantly, we will simulate the detection of fusion outcomes
and the degree of nonadiabatic transition and fermion-parity violation.
In this context, we will establish an elegant connection of
the nonadiabatic transition and fermion-parity violation
with two modification factors, while these two factors
can be determined from measuring the various occupation probabilities
of the probing quantum dot (QD), as shown in fig.\ 1.
With the help of this connection, one can extract the information
of nonadiabatic transition and fermion parity violation
caused when moving the MZMs together to fuse.

Based on the lattice model,
we will first obtain the effective coupling amplitudes of
the fusing MZMs to the probing QD and the coupling energy between themselves,
then consider near-adiabatically slow coupling during the fusion and detection stage.
The idea of slow coupling was briefly discussed in ref.\ \cite{Li24a},
in the absence of nonadiabatic transition,
where the low-energy effective Majorana mode model was employed in a direct manner.
The advantage of adiabatically switching on the coupling
is that the state evolution follows an eigenenergy state,
thus not suffering the complexity of
strong quantum oscillations of the QD occupation \cite{Li24b}.
This will drastically simplify the measurement of the QD occupation probability
and important information extraction,
in comparison with the sudden-coupling scheme analyzed in ref.\ \cite{Li24b}.
We will numerically demonstrate how the scheme can work well
by using the coarse-grained time average of
the stationary QD occupation after fusion,
in the case of slow but finite-speed fusing and probing coupling.

\vspace{0.1cm}
{\flushleft\it Theoretical Formulation}.---
As schematically shown in fig.\ 1, we will consider to fuse the pair of
Majorana modes $\gamma_2$ and $\gamma_3$,
and demonstrate the nontrivial fusion rule $\gamma_2\times \gamma_3 = I + \psi$,
which means that the fusion will stochastically
annihilate $\gamma_2$ and $\gamma_3$ into a vacuum ($I$)
or create a regular fermion ($\psi$).
For this purpose, $\gamma_2$ and $\gamma_3$
should come from different Majorana pairs with definite fermion parities.
Therefore, in practice, we need to initially prepare the Majorana pairs
$(\gamma_1,\gamma_2)$ and $(\gamma_3,\gamma_4)$
in definite fermion parity state, e.g., in state $|0_{12}0_{34}\ra$.
In this work, hereafter, we will use $n_{ij}=0$ and $1$
to denote the empty and occupied states of the regular fermion $f_{ij}$,
which is defined from the Majorana modes $\gamma_i$ and $\gamma_j$.
According to the proposal in ref.\ \cite{NC22}, this can be realized as follows.
By means of mini-gate-voltage control,
first, move $\gamma_2$ and $\gamma_3$ to the ends of the two wires,
close to $\gamma_1$ and $\gamma_4$;
then, empty the occupations of the regular fermions $f_{12}$ and $f_{34}$.
Starting with $|0_{12}0_{34}\ra$,
move $\gamma_2$ and $\gamma_3$ back to the central part to fuse.

In this work, we assume that the TSC quantum wire shown in fig.\ 1
is realized by a semiconductor nanowire
proximity-coupled to an $s$-wave superconductor.
Following ref.\ \cite{Ali18}, the lattice model Hamiltonian reads as
\bea\label{H-QW}
&& H_{\rm QW} = -\frac{W}{2} \sum_{j \sigma}(c_{j, \sigma}^{\dagger} c_{j+1, \sigma}
+c_{j+1, \sigma}^{\dagger} c_{j, \sigma})    \nl
&& +\sum_{j \sigma} (W-\mu_j) c_{j, \sigma}^{\dagger} c_{j, \sigma}
+V_{z} \sum_{j \sigma \sigma^{\prime}} c_{j, \sigma}^{\dagger}
(\sigma^{z})_{\sigma \sigma^{\prime}} c_{j, \sigma^{\prime}}    \nl
&& +\frac{\alpha_{\rm so}}{2} \sum_{j \sigma \sigma^{\prime}}
\left[ c_{j, \sigma}^{\dagger}(i \sigma^{y})_{\sigma \sigma^{\prime}}
c_{j+1, \sigma^{\prime}}+c_{j+1, \sigma^{\prime}}^{\dagger}
(i \sigma^{y})_{\sigma \sigma^{\prime}} c_{j, \sigma} \right]    \nl
&& +\Delta \sum_{j}(c_{j, \uparrow} c_{j, \downarrow}
+c_{j,\downarrow}^{\dagger} c_{j, \uparrow}^{\dagger})  \,.
\eea
In this Hamiltonian,
$c_{j \sigma}$ is the electron annihilation operator
at the $j$th lattice site and with spin $\sigma$,
$W$ is the hopping energy between two nearest sites,
$\mu_j$ is the local chemical potential
which can be modulated by electrical mini-gates,
$V_z$ represents the Zeeman energy,
$\alpha_{\rm so}$ is the Rashba spin-orbit coupling energy,
and $\Delta$ is the proximity-effect-induced superconducting energy gap.
Moreover, the Pauli matrices $\sigma^z$ and $\sigma^y$ are introduced.
For the purpose of numerically simulating the motion of the MZMs,
we need to apply the Bogoliubov de-Gennes (BdG) formalism.
The BdG Hamiltonian matrix $H_{\rm BdG}$ can be constructed through the identity
$H_{\rm QW} = \frac{1}{2} \hat{\Psi}^{\dagger} H_{\rm BdG} \hat{\Psi}$,
with
$\hat{\Psi}=(c_{1\uparrow}\cdots c_{N\uparrow},c_{1\downarrow}
\cdots c_{N\downarrow},c_{1\uparrow}^{\dagger}\cdots c_{N\uparrow}^{\dagger},
c_{1\downarrow}^{\dagger}\cdots c_{N\downarrow}^{\dagger})^{T}$,
the so-called Nambu spinor.

Moving the MZMs can be realized via the control of mini-gate voltages
as proposed in ref.\ \cite{NC22},
to sequentially make transitions from non-topological to topological regime.
After moving $\gamma_2$ and $\gamma_3$ to the central part, as shown in fig.\ 1,
they will be fused and the fusion outcomes will be probed by a quantum dot
which is monitored by a charge sensitive point-contact (PC) detector.
In this work, rather than considering a fast/sudden coupling with the QD,
as in ref.\ \cite{Li24b},
we will consider a near-adiabatic slow coupling scheme.
This scheme has the advantage of avoiding
the complexity of strong charge oscillations of the QD occupation.
In the slowly coupling case, the evolution just follows the instantaneous eigenstate
of the coupled subsystem of the $(\gamma_2,\gamma_3)$ pair and the QD.
In the final state after fusion, the QD has a stationary occupation,
which would benefit its measurement by the nearby PC detector.

During moving the Majorana pair $\gamma_2$ and $\gamma_3$
from the ends to the central part, before slowly coupling to the QD,
nonadiabatic transition and even violation of the initial fermion parity
may take place, owing to practical restrictions
that require the moving speed not too slow.
In order to simulate these effects, the moving dynamics can be simulated
by solving the time-dependent BdG (TDBdG) equation
$i \partial_t  |\Psi\ra = H_{\rm BdG} |\Psi\ra$,
in the lattice electron-hole state basis.
Actually, the BdG Hamiltonian matrix $H_{\rm BdG}$
can be also understood as constructed
under the basis of electron and hole states of the lattice sites,
i.e., $\{|e_{\xi,j\sigma}\ra,|h_{\xi,j\sigma}\ra; j=1,2,\cdots, N\}$.
Here, the index $\xi=L,R$ denotes the left and right TSC wires.
Accordingly, the wavefunction of the TSC wire can be expressed as
$|\Psi_{\xi}(t)\ra=\sum_{j,\sigma} (u_{\xi,j\sigma}(t)|e_{\xi,j\sigma}\ra
+v_{\xi,j\sigma}(t)|h_{\xi,j\sigma}\ra)$.
To identify the low-energy states, from which the MZMs are defined,
and characterize nonadiabatic transitions to excited Bogoliubov quasiparticle states,
we may recast the wire state as
\bea\label{wire-WF}
&& |\Psi_{\xi}(t)\ra = \alpha_{\xi} |\psi_{\xi,-E_0}\ra + \beta_{\xi}
|\psi_{\xi,+E_0}\ra  \nl
&& ~~~
+\sum_{n\neq 0} (\alpha_{\xi n} |\psi_{\xi,-E_n}\ra + \beta_{\xi n} |\psi_{\xi,+E_n}\ra ),
\eea
in which the instantaneous eigenstates are obtained through
$H^{(\xi)}_{\rm BdG}(t)|\psi_{\xi,\pm E_n}(t)\ra = \pm E_n(t)|\psi_{\xi,\pm E_n}(t)\ra$.
In the BdG formalism, the negative energy state $|\psi_{\xi,-E_n}\ra$
is the charge-conjugated counterpart
of the positive energy state $|\psi_{\xi,+E_n}\ra$,
holding the `particle' and `anti-particle' corresponding relation.

After completing the first stage of moving,
let us consider $\gamma_2$ and $\gamma_3$ to start fusing
and coupling to the QD, adiabatically, with the state
\bea\label{WF-LR}
 |\Psi_{\rm LR}(0)\ra
&& = (\alpha_{\rm L}|0_{12}\ra + \beta_{\rm L}|1_{12}\ra)   \nl
&& ~\otimes(\alpha_{\rm R}|0_{34}\ra + \beta_{\rm R}|1_{34}\ra) \,.
\eea

Here, we assumed the Majorana fusion and coupling to the QD
dominantly taking place within the subspace of low-energy states,
with those high-energy states being gapped out
from the full wire states of the above \Eq{wire-WF}.
In \Eq{WF-LR}, we also converted the positive and negative energy states
in the BdG formalism into occupation number states.
Notice that the initial states of the two wires
are $|0_{12}\ra$ and $|0_{34}\ra$.
The presence of $|1_{12}\ra$ and $|1_{34}\ra$
in each {\it isolated} wires
indicates the feature of fermion parity breaking,
while the reduction of $|\alpha_{\rm L}|^2$ and $|\alpha_{\rm R}|^2$
is associated with the nonadiabatic transition (to the excited quasiparticle states).

In order to reveal the charge transfer
between the fused Majorana modes and the QD more clearly,
using the basis $\{|n_{14} n_{23}\ra\}$, we reexpress the state as
\bea\label{tot-0}
|\Psi_{\rm LR}(0)\ra
&=& \frac{1}{\sqrt{2}} (A_{+}|0_{14}\ra + B_{+}|1_{14}\ra )|0_{23}\ra  \nl
&&+ \frac{i}{\sqrt{2}} (A_{-}|1_{14}\ra + B_{-}|0_{14}\ra )|1_{23}\ra  \,.
\eea
Here we introduced the notation
$A_{\pm}=\alpha_{\rm L}\alpha_{\rm R} \pm \beta_{\rm L}\beta_{\rm R}$
and $B_{\pm}=\alpha_{\rm L}\beta_{\rm R} \pm \beta_{\rm L}\alpha_{\rm R}$.
Following ref.\ \cite{NC22}, we consider to couple
a single-level QD (with energy $\epsilon_{\rm D}$)
to the fused MZMs $\gamma_2$ and $\gamma_3$,
with coupling amplitudes $\lambda_2$ and $\lambda_3$,
which can be determined from the lattice model, as numerically shown in fig.\ 2.
In terms of the regular fermion $f_{23}$, the coupling Hamiltonian reads as
$H' = (\lambda_{\rm N} d^{\dagger} f_{23}
+ \lambda_{\rm A} d^{\dagger} f_{23}^{\dagger})  + {\rm h.c.}$,
where $d^{\dagger}$ is the creation operator of the QD electron.
The two coupling amplitudes are given by
$\lambda_{\rm N,A}=\lambda_2 \pm i \lambda_3$.
$\lambda_{\rm N}$ corresponds to the usual normal tunneling process
(N-channel, associated with $|1_{23}\ra$),
while $\lambda_{\rm A}$ the Andreev process
(A-channel, associated with $|0_{23}\ra$) owing to Cooper pair splitting.

We assume the QD initially prepared in an empty state, $|0_{\rm d}\ra$.
Then, the probe-coupling evolution is given by
$|\Psi_{\rm tot}(t)\ra=U_{\rm c}(t)(|\Psi_{\rm LR}(0)\ra \otimes |0_{\rm d}\ra)$.
For either the N-channel, or the A-channel,
the adiabatic coupling evolution would follow one of
the instantaneous eigenstates in each channel,
associated with the initial state $|1_{23}0_{\rm d}\ra$ or $|0_{23}0_{\rm d}\ra$,
when adiabatically switching on the coupling $\lambda_{\rm N}$ or $\lambda_{\rm A}$,
as schematically shown in fig.\ 2.
Formally, the two instantaneous eigenstates
evolution-tracked in the both channels can be expressed as
\bea\label{Uct}
U_{\rm c}(t) |0_{23}0_{\rm d}\ra
&=& \alpha_{\rm A}(t) |0_{23}0_{\rm d}\ra +  \beta_{\rm A}(t) |1_{23}1_{\rm d}  \ra  \,, \nl
U_{\rm c}(t) |1_{23}0_{\rm d}\ra
&=& \alpha_{\rm N}(t) |1_{23}0_{\rm d}\ra +  \beta_{\rm N}(t) |0_{23}1_{\rm d}  \ra  \,,
\eea
while the superposition coefficients
can be analytically obtained in the adiabatic limit.

\begin{figure}
  \centering
  \includegraphics[scale=0.35]{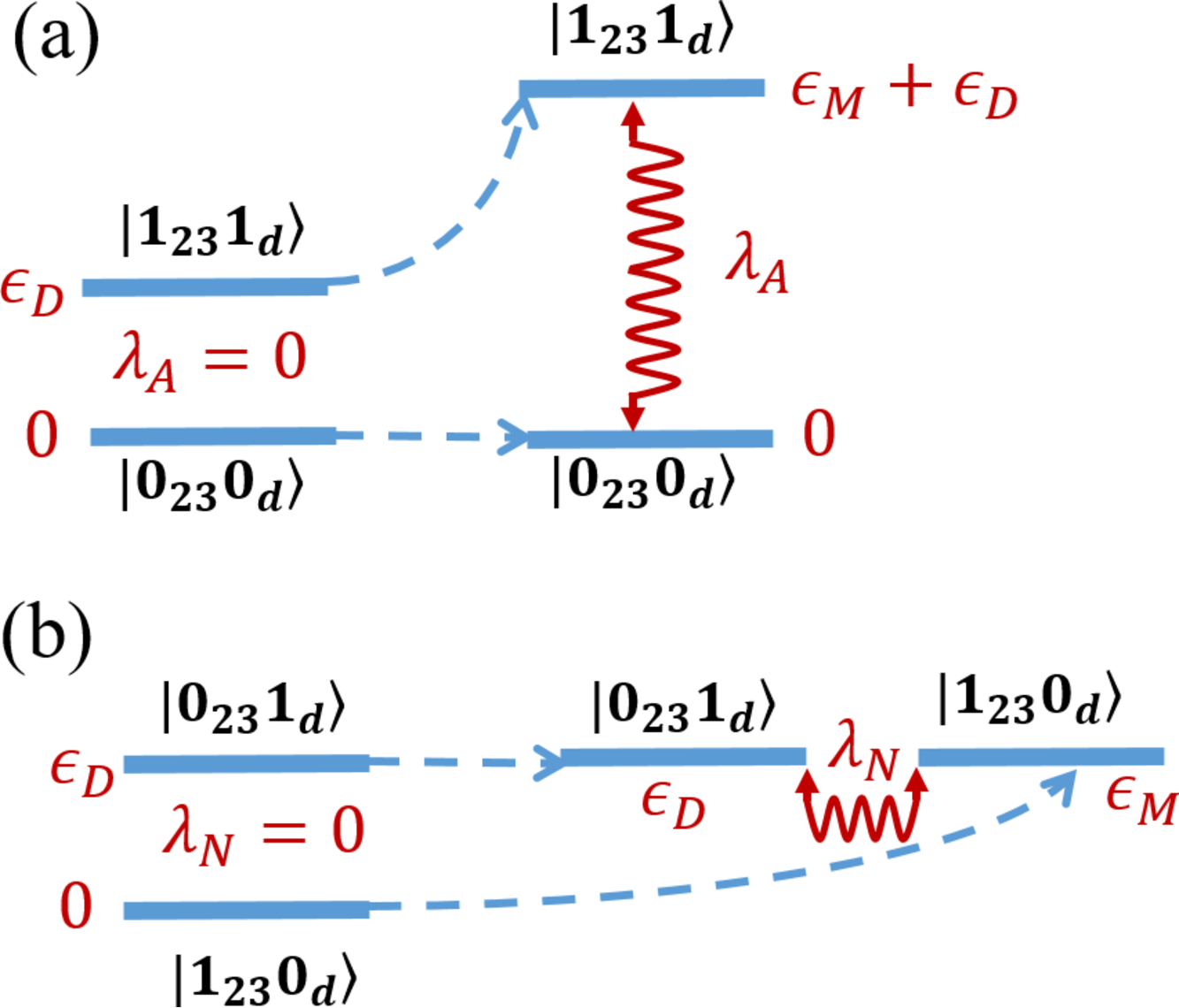}
  \caption{
Energy diagrams for adiabatically switching on the coupling
of $\gamma_2$ and $\gamma_3$ to the QD, and between themselves.
The evolution (starting from zero coupling)
is indicated by the dashed arrow lines.
In adiabatic limit, the evolution follows the instantaneous eigenstates
associated with the initial states $|0_{12}0_d\ra$ and $|1_{12}0_d\ra$,
which correspond to the fusion outcomes $I$ and $\psi$, respectively.
In this coupling scheme,
the QD occupation does not suffer strong oscillations,
benefiting thus its readout by the PC detector as shown in fig.\ 1.      }
\end{figure}

Straightforwardly, the QD occupation can be obtained as
\bea\label{Pdt}
&& P_{\rm d}(t) = \la \Psi_{\rm tot}(t)|d^{\dagger} d
           |\Psi_{\rm tot}(t)\ra   \nl
&& ~ = \frac{1}{2}\left( M_{\rm A}P_{\rm d}^{(\rm A)}(t)
+  M_{\rm N}P_{\rm d}^{(\rm N)}(t) \right) \,.
\eea
Here we introduced $P_{\rm d}^{(\rm A,N)}=|\beta_{\rm A,N}|^2$
and $M_{\rm A,N} = |A_{\pm}|^2+|B_{\pm}|^2$.
Assuming that $\epsilon_D=\epsilon_M$,
i.e., the dot level equals the energy of the fused regular fermion $f_{23}$,
we have $P_{\rm d}^{(\rm N)}=1/2$.
In the adiabatic limit, one can also find
\bea
P^{(\rm A)}_{\rm d} = |\beta_A|^2
=\frac{(\Omega_A-\epsilon_M)^2}{(\Omega_A-\epsilon_M)^2 + 2|\lambda|^2} \,,
\eea
where $\Omega_A = \sqrt{\epsilon^2_M + 2|\lambda|^2}$,
under the conditions $\epsilon_D=\epsilon_M$ and $\lambda_2=\lambda_3=\lambda$.
In the absence of nonadiabatic transition, the overall occupation probability
of an electron in the QD is simply
$P_{\rm d}= (P^{(\rm N)}_{\rm d}+P^{(\rm A)}_{\rm d})/2$.
However, in the presence of nonadiabatic transition,
the factors $M_{\rm A}$ and $M_{\rm N}$ in the above \Eq{Pdt}
play a role of {\it modification} to the result in ideal case.

Very importantly, one can check the following relation
\bea\label{P23}
&& M_{\rm N}-M_{\rm A}= 2 \la P_{23}\ra  \nl
&& = -4\, {\rm Re}[(\alpha_{\rm L}\alpha_{\rm R})^*(\beta_{\rm L}\beta_{\rm R})
        + (\alpha_{\rm L}\beta_{\rm R})^*(\beta_{\rm L}\alpha_{\rm R})] \,.
\eea
Here, $\la P_{23}\ra$ denotes the quantum average
of the parity operator $P_{23}=i\gamma_2\gamma_3$
over the state $|\Psi_{\rm LR}(0)\ra$, given by \Eq{WF-LR} or (\ref{tot-0}).
This elegant result indicates that we can infer
the deviation from the statistics of outcomes of nontrivial fusion
in ideal case, which requires $\la P_{23}\ra=0$.
That is, for any (initial) state $|n_{12}n_{34}\ra$ with definite fermion parity,
the fusion of $\gamma_2$ and $\gamma_3$ will result in
the statistical average
$\la P_{23}\ra = 0$, as formally proved in ref.\ \cite{BNK20}.
Actually, this result reflects the key feature of nontrivial fusion
with equal weight outcomes of $I$ and $\psi$.
From the above result, we find also that nonzero $\la P_{23}\ra$ is caused
by the mixing of $|1_{12}\ra$ and $|1_{34}\ra$
with the initially prepared states $|0_{12}\ra$ and $|0_{34}\ra$,
thus violating the definite fermion parity condition of each wire.
In other words, the nonadiabatic transition would induce
a breaking of the initial fermion parity associated with the MZMs.

Also, in the symmetric case, say,
$|\alpha_{\rm L}|=|\alpha_{\rm R}|=|\alpha_0|$
and $|\beta_{\rm L}|=|\beta_{\rm R}|=|\beta_0|$,
we find another important relation
\bea\label{Pex2}
M_{\rm N}+M_{\rm A} = 2 (|\alpha_0|^2+|\beta_0|^2)^2 \,.
\eea
Using this formula, one can infer the nonadiabatic transition probability
in each wire, i.e., $P_{\rm ex}=1-(|\alpha_0|^2+|\beta_0|^2)$.
Therefore, as shown by \Eqs{P23} and (\ref{Pex2}),
we established an important connection of $\la P_{23} \ra$ and $P_{\rm ex}$
with the two modification factors $M_{\rm N}$ and $M_{\rm A}$,
while $M_{\rm N}$ and $M_{\rm A}$ can be determined via \Eq{Pdt}
from measuring the various occupation probabilities
of the probing QD, as conceptually shown in fig.\ 1
and will be numerically demonstrated in the following by the results of fig.\ 6.

\vspace{0.1cm}
{\flushleft\it Results and Discussion}.---
Following ref.\ \cite{Ali18}, in the numerical simulations of this work,
we choose the wire parameters (in a reduced arbitrary system of units) as:
the hopping energy $W=1$, the superconducing gap $\Delta=0.3$,
the Zeeman energy $V_{\rm z}=0.4$,
the spin-orbit-interaction (SOI) strength $\alpha_{\rm so}=0.3$,
and the chemical potentials
$\mu_j=\mu_{\rm T}=0$ and $\mu_j=\mu_{\rm nT}=-0.45$
for the topological and non-topological phases.

\begin{figure}
  \centering
  \includegraphics[scale=0.6]{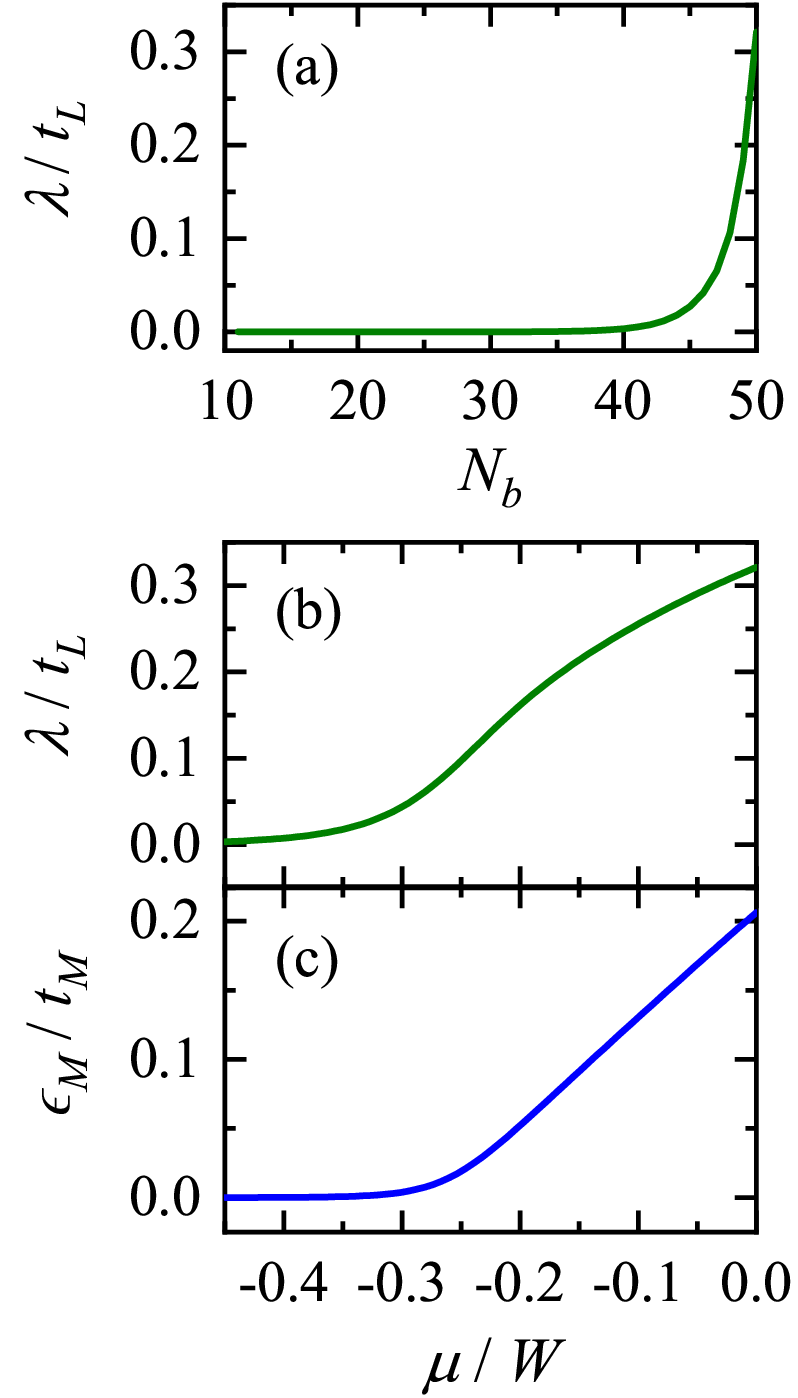}
  \caption{
Effective coupling strength $\lambda$ of the Majorana mode (e.g. $\gamma_2$)
to the probing QD, in (a) and (b),
and the coupling energy $\epsilon_{\rm M}$ between $\gamma_2$ and $\gamma_3$, in (c),
when the Majorana modes are transported close to the QD and to each other.
$N_{\rm b}$ denotes the boundary
between the topological and non-topological segments.
In (b) and (c), we assume $N_{\rm b}=40$ and
modulating the chemical potential of the last segment of 10 lattice sites.   }
\end{figure}

\begin{figure}
  \centering
  \includegraphics[scale=0.5]{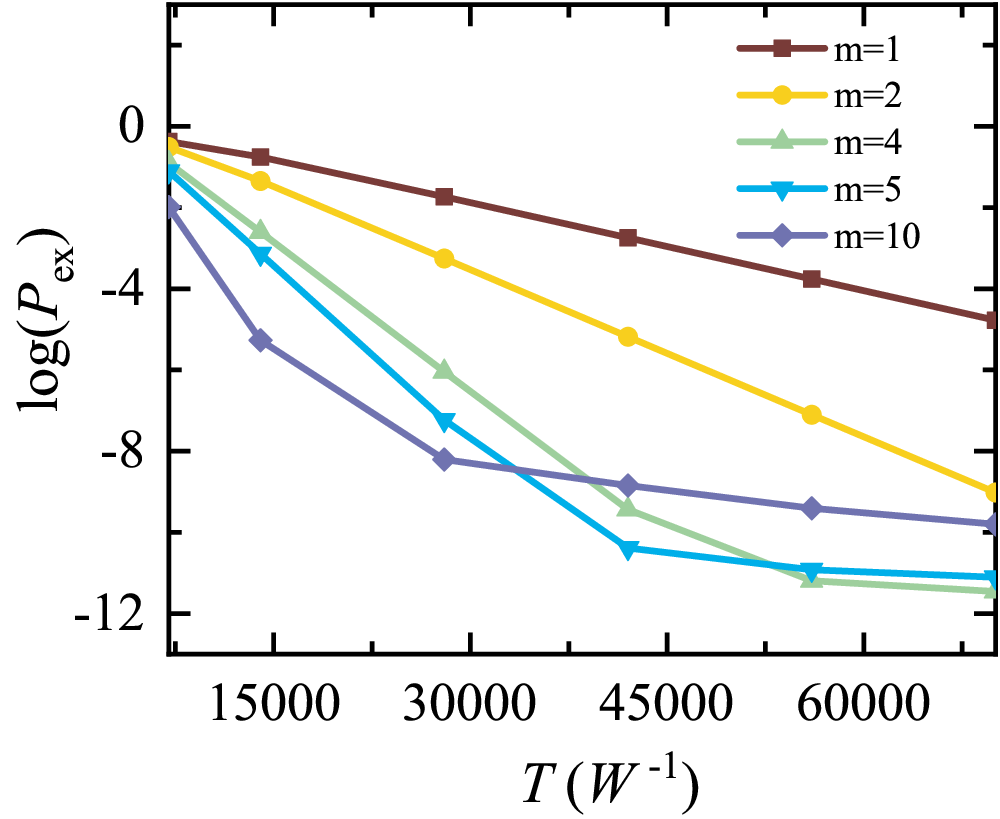}
  \caption{
Nonadiabatic transition probability {\it versus} the total transport time $T$
of different multi-segments modulation schemes.
We consider moving $\gamma_2$
by modulating the topological and non-topological phases boundary
from the $10_{\rm th}$ to $50_{\rm th}$ lattice site.
For an $m$-segments modulation scheme,
the time of moving the zero mode through each segment is $\tau_m=T/m$.    }
\end{figure}

In fig.\ 3, based on the lattice model \Eq{H-QW}
and the end-lattice-site coupling parameters
$t_{\rm L}$, $t_{\rm R}$, and $t_{\rm M}$ as schematically shown in fig.\ 1,
we show the effective coupling strength $\lambda$ of the Majorana mode (e.g. $\gamma_2$)
with the probing QD, in fig.\ 3(a) and (b),
and the coupling energy $\epsilon_{\rm M}$ between $\gamma_2$ and $\gamma_3$, in fig.\ 3(c),
as the Majorana modes are transported close to the QD and to each other.
In fig.\ 3(a), we display the change of the coupling strength $\lambda$
with the variation of the boundary between
the topological and non-topological segments,
which is located at the $N_{\rm b}$-th site (denoted in this plot).
In fig.\ 3(b) and (c),
we illustrate the gradual establishment of $\lambda$ and $\epsilon_{\rm M}$,
by considering the boundary at $N_{\rm b}=40$
and modulating the chemical potential of the last segment of 10 lattice sites,
which causes the transition from non-topological to topological phase.
Actually, this is the modulation scheme of
near-adiabatic coupling to be studied later in this work.

Compared to the slowly coupling of $\gamma_2$ and $\gamma_3$ with the QD,
moving them from the sides to the central part can be faster,
owing to the larger energy scale defined by the energy gap.
It should be instructive
to display the behavior of nonadiabatic transition
with the increase of the moving speed,
and, for different modulation schemes.
In fig.\ 4, taking the left single wire as an example,
we consider moving $\gamma_2$ from $N_{\rm b}=10$ to 50 (to the right side of the wire).
In particular, we show the different nonadiabatic transitions for different modulation schemes.
That is, we consider moving $\gamma_2$ by multiple segments modulation,
via the control of mini-gates as proposed in ref.\ \cite{NC22},
to sequentially realize transitions from non-topological to topological phase.
Specifically, we consider to modulate
the chemical potential of the $J_{\rm th}$ segment according to
$\mu_J(t)=\{1-f[(t-t_J)/\tau]\}\mu_{\rm nT} + f[(t-t_J)/\tau]\mu_{\rm T}$,
with $f(s)$ a monotonically increasing function
and satisfying $f(0)=0$ and $f(1)=1$.
Following ref.\ \cite{Ali18}, we assume $f(s)=\sin^2(s\pi/2)$.
For an $m$-segments modulation scheme, with total time $T$,
the time of moving the MZM through each segment
is $t_{J+1}-t_J=T/m\equiv \tau_m$.

From the results displayed in fig.\ 4,
we find that, in the regime of small $T$ (relatively fast transport),
the nonadiabatic transition becomes weaker
with increasing the segment number $m$ of sequential modulation.
However, in the regime of larger $T$ (relatively slow transport),
we notice crossing behavior of the $P_{\rm ex}$-{\it versus}-$T$ curves,
e.g., the $m=10$ curve with those of $m=5$ and $4$, respectively,
and the $m=5$ and $4$ curves themselves.
This crossing behavior indicates that the nonadiabatic transition increases
with the increase of $m$, after the crossing point.
This actually indicates that, with the increase of $T$,
there exists an optimal $m$-segments modulation scheme.
More careful investigation and analysis for this interesting behavior
will be performed and reported elsewhere, in a separate work.

\begin{figure}
  \centering
  \includegraphics[scale=0.4]{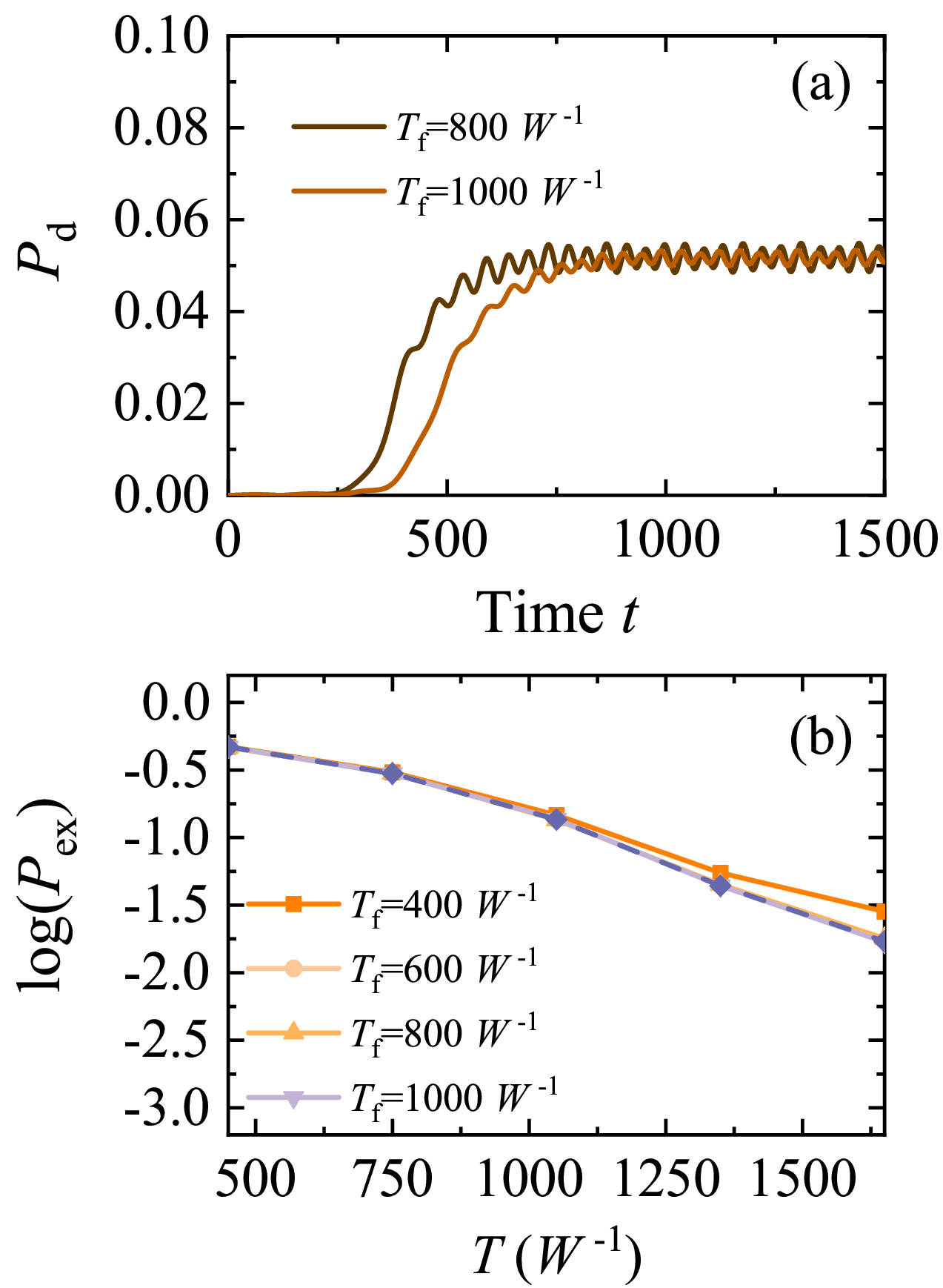}
  \caption{
(a) Occupation probability of the QD
when probing the fusion outcomes of $\gamma_2$ and $\gamma_3$,
applying the (3+1)-segments modulation scheme.
The first 3-segments ($3\times 10$ lattice sites) modulation
takes time $T=450~W^{-1}$,
while the last-segment (10 lattice sites) modulation
takes time $T_f=800~W^{-1}$ and $1000~W^{-1}$, respectively.
Owing to finite $T_f$, after fusion, the QD occupation suffers
slight oscillations with, however, convergent mean value of the adiabatic limit.
(b) For the same (3+1)-segments modulation scheme, additional information
for the negligible influence of the last-segment modulation (with time $T_f$)
on the result of nonadiabatic transition.
The $T$ dependence shows the nonadiabatic transition behavior
of the first 3-segments modulation (using time $T$).
In the simulation, we have assumed $\epsilon_{\rm D}=\epsilon_{\rm M}=1.5\lambda$.  }
\end{figure}

\begin{figure}
  \centering
  \includegraphics[scale=0.4]{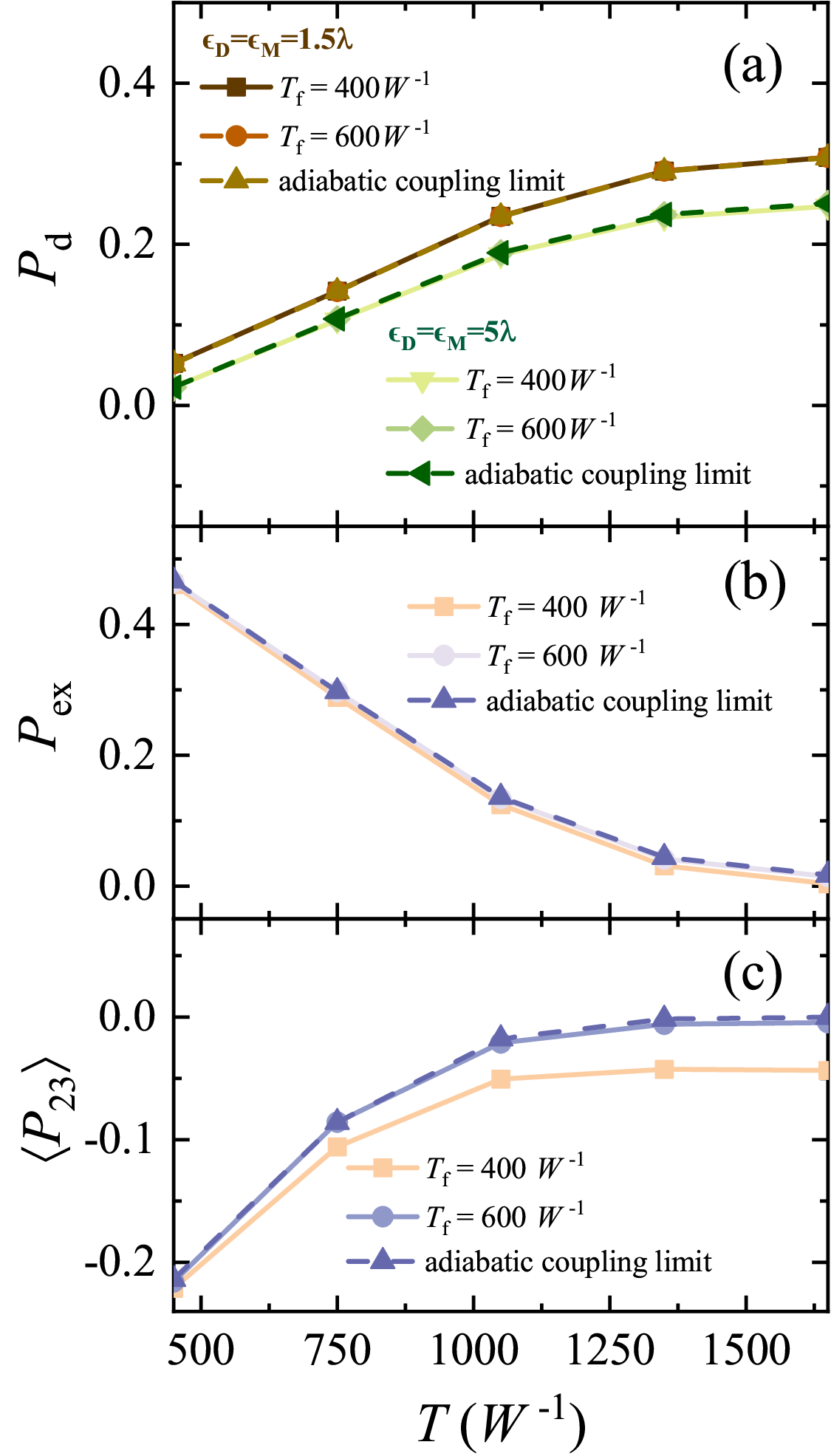}
  \caption{
Setting $\epsilon_{\rm D}=\epsilon_{\rm M}=1.5\lambda$ and $5\lambda$,
the simulated QD occupation probabilities, shown in (a),
are used to infer the factors $M_{\rm A}$ and $M_{\rm N}$ by means of \Eq{MAMN},
and further to determine $P_{\rm ex}$ in (b), and $\la P_{23} \ra$ in (c).
In (b) and (c), the results from $T_f=600~W^{-1}$ and $800~W^{-1}$
are compared with that under adiabatic coupling limit,
showing a good agreement.
In the simulation, the same (3+1)-segments modulation scheme as in fig.\ 5 is considered.  }
\end{figure}

In the following, we display simulated results
of QD probing by near-adiabatic coupling for the fusion outcomes,
in the presence of nonadiabatic transition during the transporting stage of MZMS.
In the simulation, we assume the QD level in resonance with the coupling energy
of $\gamma_2$ and $\gamma_3$ after fusion ($\epsilon_{\rm D}=\epsilon_{\rm M}$),
and symmetric coupling to the QD, $\lambda_2=\lambda_3=\lambda$.
From the results in fig.\ 3(b) and (c),
we know that $\epsilon_{\rm M}=0.2 t_{\rm M}$
and $\lambda=0.32 t_{\rm L}$, after the fusion completed.
We set $t_{\rm M}=0.25$, thus $\epsilon_{\rm D}=\epsilon_{\rm M}=0.05$.
We also assume to modulate the coupling strengths $t_{\rm L}$ and $t_{\rm R}$
(notice that $t_{\rm L}=t_{\rm R}$),
such that $\epsilon_{\rm D}=\epsilon_{\rm M}=1.5\lambda$,
for the results in fig.\ 5,
and adding one more choice of $5\lambda$ in fig.\ 6,
for the need of important information extraction.
In fig.\ 5(a) we show the QD occupation
from the $(m+1)$-segments-modulation scheme,
which is performed symmetrically for each of the left and right quantum wires.
Specifically, the first $m$-segments-modulation
is performed fast, within the total time $T=m\tau_m$,
while the last-segment-modulation is performed relatively slowly, within time $T_f$,
to ensure the condition of near-adiabatic coupling
of $\gamma_2$ and $\gamma_3$ to the QD, and to each other.
In the simulation of the results in fig.\ 5(a),
we specified $m=3$ and $T=450~W^{-1}$.
We display the results for $T_f=800~W^{-1}$ and $1000~W^{-1}$.
Rather than the ideally adiabatically switching on the coupling to the probing QD,
here we notice slight oscillations of the QD occupation,
from the real-time dynamics simulation.
However, it is clear that the mean value of the oscillating occupation probability
converges to the result predicted under the adiabatic limit.
In practice, as to be shown later in fig.\ 6,
one can extract the QD occupation in adiabatic limit
from the mean value of the oscillations,
to identify the nontrivial fusion of $\gamma_2$ and $\gamma_3$,
and meanwhile to infer the nonadiabatic transition and fermion-parity breaking
in the first stage of the $m$-segments modulation.
In fig.\ 5(b), we provide additional information for
the influence of the last-segment-modulation (with time $T_f$),
on the result of nonadiabatic transition probability
in the first stage of the 3-segments-modulation (taking time $T$).
We find that the modulations with $T_f=600~W^{-1}$, $800~W^{-1}$
and $1000~W^{-1}$ have negligible influences.

Further, in fig.\ 6(a), we show the results of QD occupation probability
{\it versus} the transporting time $T$ of $\gamma_2$ and $\gamma_3$,
by taking the mean value of stationary-state oscillations as shown in fig.\ 5(a).
To demonstrate the feasibility of the near-adiabatic-coupling approach,
we adopt more conservative choice by assuming
the last-segment-modulation with $T_f=400~W^{-1}$ and $600~W^{-1}$.
For the purpose of extracting the key factors
$M_{\rm A}$ and $M_{\rm N}$ from measurement data,
we assume two groups of $\epsilon_{\rm D}$ and $\epsilon_{\rm M}$,
say, $\epsilon_{\rm D}=\epsilon_{\rm M}=1.5\lambda$ and $5\lambda$.
For the both choices of $\epsilon_{\rm D}$ and $\epsilon_{\rm M}$,
since $\epsilon_{\rm D}=\epsilon_{\rm M}$,
we have $P_{\rm d}^{\rm (N)}=1/2$.
In real experiment, the condition of $\epsilon_{\rm D}=\epsilon_{\rm M}$
can be slightly relaxed, thus we may obtain different values of $P_{\rm d}^{\rm (N)}$.
We can also obtain two values of $P_{\rm d}^{\rm (A)}$,
just denoting in this context as $a$ and $b$.

Actually, in practice, we can obtain the individual
$P^{\rm (N)}_{\rm d}$ and $P^{\rm (A)}_{\rm d}$ via measurements as follows.
As initializing the empty occupation $n_{12}=0$ and $n_{34}=0$
of the regular fermions $f_{12}$ and $f_{34}$,
we can assume the same way to empty the occupation of the regular fermion $f_{23}$
associated with the fused MZMs $\gamma_2$ and $\gamma_3$,
by introducing an additional tunnel-coupled side quantum dot (electron-mediating-QD)
and modulating the dot energy level $\widetilde{\epsilon}_{\rm D} \simeq \epsilon_{\rm M}$,
while the electron-mediating-QD is tunnel-coupled to an outside reservoir
with Fermi level lower than $\widetilde{\epsilon}_{\rm D}$.
This can ensure $n_{23}=0$. Starting with it,
one can measure the stationary $P_{\rm d}^{\rm (A)}$ via the near-adiabatic coupling scheme.
In order to measure and obtain $P_{\rm d}^{\rm (N)}$,
after ensuring $n_{23}=0$ as explained above,
one can inject an electron from the reservoir into an occupation
of the regular fermion $f_{23}$ through the electron-mediating-QD,
by raising the Fermi level of the reservoir above $\widetilde{\epsilon}_{\rm D}$.
Then, starting with $n_{23}=1$,
one can measure the stationary $P_{\rm d}^{\rm (N)}$
via the near-adiabatic coupling scheme.

For each choice of $\epsilon_{\rm D}$ and $\epsilon_{\rm M}$,
the QD occupation probability $P_{\rm d}$
can be also measured by the PC detector,
with result as simulated in fig.\ 6(a).
Let us denote them as $A$ and $B$.
Thus, based on \Eq{Pdt}, we have two equalities and can solve to obtain
\bea\label{MAMN}
M_{\rm A} &=& 2(A-B) /(a-b) \,,  \nl
M_{\rm N} &=& 4(aA-bB) /(a-b)  \,.
\eea
In this way,
after knowing $M_{\rm A}$ and $M_{\rm N}$ from the measurement data,
we can make two types of inferences.
One is that the coexistence of nonzero $M_{\rm A}$ and $M_{\rm N}$
indicates the presence of two fusion outcomes, say, $I$ and $\psi$,
serving thus as the evidence of the nontrivial fusion.
Otherwise, one of $M_{\rm A}$ and $M_{\rm N}$ should be zero.
Second, from the obtained $M_{\rm A}$ and $M_{\rm N}$,
we can infer the results of $P_{\rm ex}$ and $\la P_{23}\ra$,
by applying the important relations of \Eqs{Pex2} and (\ref{P23}).
In fig.\ 6(b) and (c), we show the results
from two slow-coupling modulations,
with $T_f=400~W^{-1}$ and $600~W^{-1}$, respectively.
By comparing with the results under adiabatic coupling limit,
we find that the modulation with $T_f=400~W^{-1}$ has some deviation,
owing to the relatively fast modulation which causes some extra nonadiabatic transition,
while the modulation with $T_f=600~W^{-1}$ can give desirable results
in good agreement with the adiabatic coupling limit.

\vspace{0.1cm}
{\flushleft\it Summary and Discussion}.---
We proposed and simulated a near-adiabatic coupling scheme
for probing the nontrivial fusion of a pair of MZMs.
The advantage of this scheme is that the state evolution during the probing stage
does not suffer the complexity of strong quantum oscillations.
This can simplify the measurement of the probing QD occupation probabilities
and important information extraction from the measurement results.
Since the nontrivial fusion requires the Majorana modes to be fused
coming from different pairs with definite fermion parity,
we presented simulation results for transport and fusion of MZMs,
based on the lattice model of Rashba TSC quantum wires.
We also illustrated how to extract the information of
nonadiabatic transition and fermion parity violation
taking place when moving the MZMs to fuse, from the prepared states.
The fusion and probing scheme analyzed in this work
is expected to guide future experiments,
while the real experimental demonstration
should be a milestone progress in this challenging field.

The basic requirement of the near-adiabatic coupling is that $T_f^{-1}$
is smaller than the energy scales of $\epsilon_M$, $\epsilon_D$ and $\lambda$.
This is also the basic requirement of various QD-assisted braiding schemes of MZMs.
Moreover, an extra essential requirement is that
both the moving time $T$ and coupling time $T_f$
should be shorter than the quasi-particle poisoning time,
while that time strongly depends on external environment and coupling strength with it.
The absence of quasi-particle poisoning is important,
since it will change the fermion parity of a pair of MZMs.
Under this condition,
the first stage of Majorana transport
may involve certain nonadiabatic transition.
However, as analyzed in this work,
it does not hinder the present detection scheme of nontrivial fusion.
In this case, one can still identify the nontrivial fusion,
and thus the underlying non-Abelian statistics of MZMs.
And, importantly, one can even extract the amount of
nonadiabatic transition and fermion-parity breaking.

Therefore, the near-adiabatic-coupling of fusion and detection
will not add much burden on the restriction of timescales,
while it can greatly simplify the detection of fusion outcomes.
This is because
measuring strong quantum oscillations of the QD occupation
by a quantum-point-contact device is a challenging task in practice.
Finally, we remark that the proposed study in present work is most suitable to
the platform of planar Josephson junctions \cite{Ren19},
which are fabricated from 2DEG and in proximity contact
with the conventional $s$-wave superconductors.
In the same 2DEG platform, the QD and QPC can by fabricated and integrated
together with the Majorana quantum junctions,
as schematically shown in Fig.\ 1 and originally proposed in ref.\ \cite{NC22}.

\vspace{0.2cm}
{\flushleft\it  Data availability statement}
:All data that support the findings of this study are included within the article (and any supplementary files).


\end{CJK*}
\end{document}